\documentclass[%
 reprint,
 showkeys,
 nofootinbib,
 amsmath,amssymb,
 pra,
 longbibliography,
 floatfix,
 ]{revtex4-1}

\usepackage{graphicx}
\usepackage{epstopdf}
\usepackage{eepic}
\usepackage{color}
\usepackage{amsmath}
\usepackage{amsthm}
\usepackage{natbib} 
\usepackage{comment,xspace}
\usepackage{multirow}

\usepackage{amsmath,amssymb,amsfonts,url,amsthm,enumerate}
\usepackage{graphicx,epsf}

\usepackage{algorithm,algcompatible}
\algnewcommand\algorithmicreturn{\textbf{return}}
\algnewcommand\RETURN{\State \algorithmicreturn}%

\usepackage{dcolumn}
\usepackage{bm}

\theoremstyle{plain}
\newtheorem{theorem}{Theorem}[section]

\newcommand{\cu}[1]{#1}

\usepackage{array}
\newcolumntype{H}{>{\setbox0=\hbox\bgroup}c<{\egroup}@{}}

\newcommand{\ket}[1]{\left|#1\right\rangle}

\newcommand{\xor}{\oplus}
\renewcommand{\AA}{{\tt AA}\xspace}
\newcommand{\AS}{{\tt AS}\xspace}
\newcommand{\A}{\mathcal{A}}
\newcommand{\MWDP}{\mathrm{\mathbf{MWDP}}}
\newcommand{\EWDP}{\mathrm{\mathbf{EWDP}}}
\newcommand{\WDP}{\mathrm{\mathbf{WDP}}}
\newcommand{\EBQP}{\mathrm{\mathbf{EBQP}}}
\newcommand{\EQP}{\mathrm{\mathbf{EQP}}}
\newcommand{\pifrac}[1]{\frac{\pi}{#1}}
\newcommand{\half}{\frac{1}{2}}
\newcommand{\tO}{\tilde{O}}

\begin{document}


\title{Error reduction of quantum algorithms}


\author{Debajyoti Bera}
\email{dbera@iiitd.ac.in}
\affiliation{Indraprastha Institute of Information Technology,
Okhla Industrial Estate Ph-III, New Delhi, India 110020}
\author{Tharrmashastha P.V.}
\email{tharmasasthapv@gmail.com}
\affiliation{Indraprastha Institute of Information Technology,
Okhla Industrial Estate Ph-III, New Delhi, India 110020}

\date{\today}

\begin{abstract}
    We give a technique to reduce the error probability of quantum algorithms that determine whether its input has a specified property of interest. The standard process of reducing
this error is statistical processing of the results of multiple independent executions of an algorithm. Denoting by $\rho$ an upper bound of this probability (\textit{wlog.}, assume $\rho \le \half$), classical techniques require $O(\frac{\rho}{[(1-\rho) - \rho]^2})$ executions to reduce the error to a negligible constant.  We investigated when and how quantum algorithmic techniques like amplitude amplification and estimation may reduce the number of executions. On one hand, the former idea does not directly benefit algorithms that can err on both yes and no answers and the number of executions in the latter approach is $O(\frac{1}{(1-\rho) - \rho})$.  We propose a novel approach named as {\em Amplitude Separation} that combines both these approaches and achieves $O(\frac{1}{\sqrt{1-\rho} - \sqrt{\rho}})$ executions that betters existing approaches when the errors are high.

In the \emph{Multiple-Weight Decision Problem}, the input is an $n$-bit Boolean function $f()$ given as a black-box and the objective is to determine the number of $x$ for which $f(x)=1$, denoted as $wt(f)$, given some possible values $\{w_1, \ldots, w_k\}$ for $wt(f)$. When our technique is applied to this problem, we obtain the correct answer, maybe with a negligible error, using $O(\log_2 k \sqrt{2^n})$ calls to $f()$ that shows a quadratic speedup over classical approaches and currently known quantum algorithms.




\end{abstract}


\maketitle





\section{Introduction}
\label{sec:intro}

Many of the famous problems for which early quantum algorithms were designed are
``decision problems'', i.e., the solution of the problem requires identifying
whether an input satisfies a given property. Inputs which evoke a ``yes'' answer
are called as ``yes''-inputs and, similarly, those that evoke a ``no'' answer
are called as ``no''-inputs. Quantum algorithms being inherently
probabilistic, it is possible for such algorithms to be error-prone. An
algorithm that makes the correct decision for every input is termed as an
``exact algorithm'', otherwise the algorithm is a probabilistic one. This note
concerns probabilistic quantum algorithms and the techniques to reduce their
error. Specifically, we look at algorithms with {\em bounded non-zero errors} in the following
sense: the probability of error for yes-inputs is upper bounded by $\rho_y \in
(0,1)$ and
the probability of error for no-inputs is upper bounded by $\rho_n \in (0,1)$.
Without loss of generality, we will assume that $\rho_n \le \rho_y$, else the notion
of ``yes'' and	``no'' inputs can be interchanged; similarly, we can assume that
$\rho_y + \rho_n \le 1$, because otherwise, $(1-\rho_y) + (1-\rho_n) \le 1$ so
we can simply swap the ``yes''-``no'' answers.

Setting aside bespoke error reduction tactics, our focus is going to be
black-box techniques for reducing error that applies to
any algorithm. This is routinely done for
day-to-day classical algorithms by running them independently enough number of times and
analysing their output. For example, if $\rho_n$ is 0, then
it suffices to simply output {\tt ``yes''} if any execution outputs ``yes''
In fact
the versatile {\em amplitude amplification} (\AA) technique used in quantum
algorithms can also be used in such
cases~\cite{brassard2002quantum,bera2018cocoon}.

However, directly applying \AA is inadequate in reducing errors of
algorithms if both $\rho_y > 0$ and $\rho_n > 0$. What \AA does is
non-linearly multiplies the probability that the output state of an algorithm is
observed in a particular state (for which the algorithm outputs ``yes''). Therefore,
when both $\rho_y$ and $\rho_n$ are non-zero, there is a chance of error for
every input. No matter which state is used for amplification, one of
$\rho_y$ and $\rho_n$ will decrease but the other will increase, rendering $\AA$
ineffective.

\begin{figure*}[t]
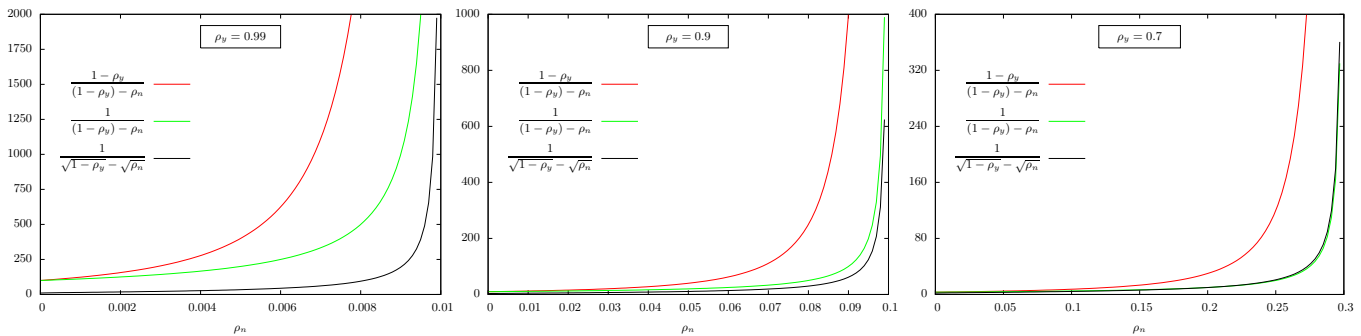

    \hspace*{-.5cm}\resizebox{0.36\linewidth}{!}{\input{distcompare-99.tex}}%
    \hspace*{-.5cm}\resizebox{0.36\linewidth}{!}{\input{distcompare-9.tex}}%
    \hspace*{-.4cm}\resizebox{0.36\linewidth}{!}{\input{distcompare-7.tex}}
    \caption{Comparison of the dominant terms in the query complexity of the
    classical, amplitude estimation based and amplitude separation based error
    reduction algorithms. The plot for $\rho_y=0.5$ (not shown) is identical to
    that of $\rho_y=0.7$ indicating that amplitude separation based algorithms are best for high
    values of $\rho_y$ and are reasonably good even for the lower
    values.\label{fig:comparison_chart}}
\end{figure*}

There are standard ``classical'' techniques for handling such algorithms.
Suppose $\A$ denotes the algorithm with error bounds $\rho_y$ and $\rho_n$.
Therefore, for a ``yes''-input, the probability of observing a ``good'' output
state would be at least $1-\rho_y$ and for a ``no''-input, the probability of
observing the same would be at most $\rho_n$. One manner in which
the error of $\A$ can be reduced (to, say, some $\delta$) is to estimate this
probability with a precision of $\pm \half[(1-\rho_y) - \rho_n]$ and
with error probability at most $\delta$. For a ``yes''-input, the estimate will be less than $\half[(1-\rho_y)
+ \rho_n]$ with probability less than $\delta$ and for a ``no''-input, the
estimate will be more than the same threshold with probability less than
$\delta$. Thus, to reduce the error of $\A$, it suffices to estimate the
probability and claim that the input is a ``yes''-input if the estimate is more
than the threshold, and a ``no''-input otherwise.
Estimating the probability
requires running $\A$ multiple times and calculating the fraction of times the
``good'' state is observed, and to achieve this within the required bounds
requires $\tO\big((1-\rho_y)/[(1-\rho_y)-\rho_n]^2\big)$ executions~\footnote{$\tilde{O}()$ hides additional insignificant $\log$-factors within $O()$.}
of $\A$.

Another possibility is the use of {\em amplitude estimation} that is a quantum
technique to estimate the probability that the output of any
algorithm is observed to be in a ``good'' state. The probability can be estimated
with any required precision --- there is a chance of error but that too can be
controlled at the expense of more operations. Use of this technique reduces the number of executions 
to $\tO(1/[(1-\rho_y)-\rho_n])$.

However, both these techniques become inefficient when $1-\rho_y
\approx \rho_n$ and both of these are small.
This note presents the {\em amplitude separation} technique, a combination of
amplitude amplification and estimation, to reduce both $\rho_y$ and $\rho_n$, even when
they are non-zero, and the number
of calls required is only $\tO(1/[\sqrt{1-\rho_y}-\sqrt{\rho_n}])$. As illustrated in
Figure~\ref{fig:comparison_chart}, this method outperforms the earlier
techniques when $1-\rho_y \to 0$ and $\rho_n \to 0$.

If the errors for all the ``yes''-inputs are same and equal to $\rho_y$, and
similarly, those for all the
``no''-inputs are equal to $\rho_n$ and if $\rho_y$ and $\rho_n$ are known
then it is possible to perform a better error reduction. Using \AA in a sophisticated
manner, Bera has shown how to obtain an algorithm that correctly
outputs {\tt ``accept''} for all ``yes''-inputs and outputs {\tt ``reject''} for
all ``no''-inputs {\em without} any probability of error (see the result that
$\EBQP=\EQP$ in \cite{bera2018cocoon}).
However, that technique crucially uses the information that all error probabilities
equal either $\rho_y$ or $\rho_n$ and are known priory --- something which we
relax in this note. Furthermore, the objective of that work was to design an
error-less method whereas we allow error, albeit tunable, as a parameter.

An immediate application of our method is an
efficient bounded-error algorithm for the {\em Multiple Weight Decision} problem
($\MWDP$). $\MWDP$ is a generalization of the
{\em Exact Weight Decision} problem ($\EWDP$) that, in turn,
generalizes the Deutsch-Jozsa's problem and the Grover's unordered search
problem~\cite{QZPRA,Braunstein07,Choi11}. The input to the $\MWDP$ problem is an $n$-bit Boolean function $f()$ given in the form
of a blackbox and a list of $k$ possible weights of $f()$: $\{0 < w_1 < w_2 < \ldots <
w_k < 2^n \}$ along with a promise that $wt(f)=w_i$ for some $i$. The weight of $f()$
is defined as $wt(f)=|\{ x \in \{0,1\}^n ~:~ f(x)=1 \}|$. The objective is to
determine the actual weight of $f()$ by making very few calls to $f()$.
$\EWDP$ can be defined as $\MWDP$ with $k=2$.

Optimal
algorithms for $\EWDP$ are known that determines the weight exactly and make
$\Theta(\sqrt{w_2(2^n-w_1)}/(w_2-w_1))$ calls~\cite{Choi11,ChoiLowerBound12} that  could be as large as
$\sqrt{2^n}$ when $w_1 \approx w_2 \ll 2^n$. Current
algorithms for $\MWDP$ give exact answer and follow two approaches~\cite{Choi11}. They either make $k-1$ calls to an $\EWDP$
algorithm, and thus, could make nearly $2^n$ calls to $f()$ (when $k \approx
\sqrt{2^n}$), or, they use a quantum counting
algorithm~\cite{brassard2002quantum} to count the number of
solutions of $f(x)=1$ but that could also
require nearly $2^n$ calls (when $wt(f) \approx 2^n$).

We use our amplitude separation technique to give an algorithm for $\MWDP$ with
{\em small error} that makes $O(\log_2 k \sqrt{2^n})$ calls to $f()$. This is
achieved by first designing a bounded-error algorithm for a variation of $\EWDP$
in which we have to determine if $wt(f) \le w_1$ or $wt(f) \ge w_2$ for given
$0 < w_1 < w_2 < 2^n$.

Our approach uses the concept
of amplitude amplification (\AA) and amplitude estimation. Even though we
describe our technique on algorithms that take its input in the form of
oracle operators, we use can a method outlined in a work by Bera to apply
\AA, and hence the technique in this note, to algorithms that is given their input $x \in
\{0,1\}^n$ in the form of an initial state $\ket{x}$ (along with ancillary
qubits in a fixed state)~\cite{bera2018cocoon}.
\section{Background}
\label{sec:amp-est}

Our method makes a subtle use of the well-known quantum amplitude estimation
algorithm so we briefly discuss the
relevant results along with the specific extension that we require.

Suppose we have an $n$-qubit quantum algorithm $\A$ that is said to ``accept'' its input when
its output qubit is observed in a specific ``good state'' upon the final
measurement. We will use $p$ to denote the probability of observing this good
state for a specific input. The value of $p$ can be estimated by purely
classical means, e.g., by running the algorithm multiple times and computing the
fraction of times the good state is observed. Amplitude estimation is a quantum
technique that essentially returns an estimate by making fewer calls to the
algorithm compared to this technique.

The estimation method uses two parameters $k$ and $m$ that we shall fix later.
The first and basic quantum amplitude estimation algorithm (say, named as $AmpEst$) was proposed by Brassard et
al.~\cite{brassard2002quantum}
that acts on two registers of $m$ and $n$ qubits, makes
$2^m$ calls to controlled-$\A$ and outputs a $\tilde{p} \in [0,1]$ that is a good
approximation of $p$ in the following sense.
\begin{theorem}
    The $AmpEst$ algorithm returns an estimate $\tilde{p}$ that has a confidence
    interval $|p-\tilde{p}| \le 2\pi k \frac{\sqrt{p(1-p)}}{2^m} +
    \pi^2 \frac{k^2}{2^{2m}}$ with probability at least $\frac{8}{\pi^2}$ if
    $k=1$ and with probability at least $1-\frac{1}{2(k-1)}$ if $k \ge 2$. If
    $p=0$ or 1 then $\tilde{p}=p$ with certainty.
\end{theorem}


The $AmpEst$ algorithm can be used to estimate $p$ with desired accuracy (at
least $3/4$) and error.
We now present an extension to the above Theorem to obtain an estimation with an
{\em additive error}, say denoted by $\epsilon$, that is at most $1/4$. We will use $\delta$ to denote the
maximum permissible error. For obtaining such an estimation, we will run 
$AmpEst$ presented above using $k=1$ and  $m$ such that $2^m = \lceil
\frac{3\pi}{2\epsilon} \rceil$. $AmpEst$ will be run
$7\left(\ln\frac{1}{\delta}\right)^{1/3} = \Theta(\ln\frac{1}{\delta})$ times to
obtain that many estimates of $p$ and the median of these obtained
    estimates is then returned as $\tilde{p}$. The total number of calls to
controlled-$\A$ is, therefore, $O(\frac{1}{\epsilon}\ln\frac{1}{\delta})$. Next
we analyse the accuracy of $\tilde{p}$.

Since $p(1-p) \le 1/4$ for any $p$, $\half e^{2\epsilon} \ge \sqrt{\frac{1}{4} +
\epsilon}$ ($\because \epsilon \le 1/4$), and $3 \ge 1+e^{2\epsilon}$, it can be
shown that $\frac{3\pi}{2\epsilon} \ge \frac{\pi}{\epsilon} \left[
\sqrt{p(1-p)} + \sqrt{p(1-p) + \epsilon} \right]$ and for $2^m \ge
\frac{3\pi}{2\epsilon}$, it can be further shown that
$2\pi\frac{\sqrt{p(1-p)}}{2^m} + \frac{\pi^2}{2^{2m}} \le
\epsilon$. Therefore, for the setting of parameters specified above,
using the above Theorem we obtain an estimate $\tilde{p}$ in each run of $AmpEst$ such that
\mbox{$\Pr[|p-\tilde{p}| \ge \epsilon] \le \delta$} with probability of error at
most $1-\frac{8}{\pi^2}$ which means the median of any number of such
estimates also satisfies the same upper-bound on its additive error. The overall
error can be reduced to any desired $\delta$ by taking a median of
$\Theta(\ln\frac{1}{\delta})$ estimates and this is a standard error reduction
technique whose proof uses Chernoff bounds.

So, to summarize this section, we have explained a method that returns an estimate
$\tilde{p}$ to the success probability $p$ of a quantum algorithm $\A$ such that
\mbox{$\tilde{p}-\epsilon \le p \le \tilde{p} + \epsilon$} with a probability at least
$1-\frac{1}{\delta}$. The method makes altogether
$O(\frac{1}{\epsilon}\ln\frac{1}{\delta})$ calls to $\A$.

\section{Amplitude Separation Algorithm}\label{sec:as}

Now we introduce the {\em Amplitude Separation} (\AS) problem and describe an
algorithm that is going to be our main technical tool.
Suppose we are given a quantum algorithm $\A$ for a decision problem;
without loss of generality, we can assume that the algorithm outputs ``yes'' if
the output qubit is observed in the state $\ket{1}$ and ``no'' if the observed
state is $\ket{0}$. Let $p$ denote the probability of observing the output qubit
in the state $\ket{1}$.  Suppose it is also given that for ``yes''-inputs $p \ge t$ and for
``no''-inputs $p \le t'$ for given $0 < t' < t < 1$. The \AS problem is to determine whether a given input is a ``yes''-input
or a ``no''-input by making black-box calls to $\A$.

There are, of course, several alternative strategies.
Consider the completely classical method of making multiple observations of $\A$
and deciding based on the number of times the output qubit is observed in the
state $\ket{1}$ -- the number of required queries to $\A$ can be obtained using
probabilistic techniques (involving Chernoff bound) and scales as
$O(\frac{1}{t-t'})$.
Another possibility would have been to use the quantum
amplitude estimation methods.
They come in various flavours and a quick summary of the relevant ones
are presented in Section~\ref{sec:amp-est}. If we use the additive-accuracy
estimation, then too the number of queries scales as in the previous
case. One can also design an estimator with a relative-accuracy but to
obtain an upper-bound on the number of queries, one would require a lower bound
on $p$ which need not be known.

The decision algorithm is presented in Algorithm~\ref{algo:decision_tau}.
For the simplicity of analysis, we use a {\em separation} variable $\beta$
chosen such that $t' = \beta^2 t$.
On a high level, our algorithm
first amplifies the amplitude of $\ket{1}$ state of the output qubit and only
after that applies amplitude estimation
since amplified probabilities have a larger gap and, therefore, are easier to
distinguish.
Recall that applying \AA $k_i$ times increases the corresponding probability
from any $\sin^2\theta$ to $\sin^2[(2k_i+1)\theta]$. 
We will see below how this allows us to solve the problem with a number of
queries to $\A$ that scales as $O(\frac{1}{\sqrt{t}})$. 
For amplitude estimation we use the additive-accuracy estimator with
additive-error $\epsilon'$ and error $\delta'$ that is explained in
Section~\ref{sec:amp-est}.

\begin{algorithm}[H]
    \caption{Amplitude Separation($\A$) \label{algo:decision_tau}}
	{\bf Parameter:} $0 < t' < t \le 1$ (thresholds)\\
	{\bf Parameter:} $\delta$ (error)\\
	{\bf Denote:} $\ket{in}$ as the initial state of $\A$
    \begin{algorithmic}[1]
	\STATE Set $\beta = \sqrt{t/t'}$, $\tau = sin^{-1}\sqrt{t}$, $s =
	\lfloor \log_3 \frac{\pi}{4\tau}\rfloor$, $\delta' = \frac{\delta}{(1+s)}$.
	\STATE Set $\epsilon' = \half(\sin^2 3^s\tau - \sin^2 3^s\beta\tau)$
	\STATE Set $\epsilon^* = \half(\sin^2 3^s\tau + \sin^2 3^s\beta\tau)$.
	\FOR {$i=0$ to $s$}
	\STATE Set $k_i = \frac{1}{2}(3^i-1)$.
	\STATE $\ket{\phi} \leftarrow $ apply amplitude amplification $k_i$
	times to $A\ket{in}$
	\STATE $\tilde{p} \leftarrow $ estimate probability of
	observing the output qubit of $\ket{\phi}$ in the state $\ket{1}$
	using ``Amplitude Estimation with additive error $\epsilon'$ and error
	$\delta'$''
	\STATE If $\tilde{p} \ge \epsilon^*$ : \RETURN ``{\tt
		accept}'' ({\it i.e.,} claim that $p \ge t$).
	\ENDFOR
	\RETURN ``{\tt reject}'' ({\it i.e.,} claim that $p \le t' = \beta^2 t$)
    \end{algorithmic}
\end{algorithm}

Now we explain how Algorithm~\ref{algo:decision_tau}
makes $\tilde{O}\left(\frac{1}{\sqrt{t}-\sqrt{t'}}\log\frac{1}{\delta}\right)$
calls to $\A$ (and $\A^\dagger$) and with probability of error at most $\delta$
returns {\tt accept} if $p \ge t$ or returns 
{\tt reject} if $p \le t'$.

To explain the claim we will use the two following trigonometric facts:
(1) for any $a < 1$ and $t \le \pi/2 $, $\sin\theta \le a$ $\sin t$ implies $\theta
\le a t$, and (2) for any $a < 1$ and $t \le \pi/4$, $a \sin t \le \sin at \le
\sqrt{a}\sin t$ (proof of these are included in Appendix~\ref{appen:lemma-proof}).

Consider $\theta \in [0,\pifrac{2}]$ such that $p = \sin^2 \theta$ and $\tau \in
[0,\pifrac{2}]$ such that $t = \sin^2 \tau$.
Then the two cases of $\theta$ that are under consideration would be (i) $\sin \theta \ge \sin \tau$ and (ii) $\sin \theta \le \beta \sin \tau$.
Following a common technique of analysing amplitude amplification
techniques~\cite{Chakraborty2016}, it will be helpful to break the interval
[$\tau, \frac{\pi}{2}$] into these intervals:
	\begin{align*}
	    R_0 & = \left[\psi, \frac{\pi}{2}\right], R_1 =
	    \left[\frac{1}{3}\psi, \psi\right], R_2 = \left[\frac{1}{3^2}\psi,
	    \frac{1}{3}\psi\right] \\
	    \ldots R_i & = \left[\frac{1}{3^i}\psi, \frac{1}{3^{i-1}}\psi\right]
	    \ldots R_s = \left[\tau=\frac{1}{3^s}\psi, \frac{1}{3^{s-1}}\psi\right]
	\end{align*}
	where $\psi = 3^s\tau$ and $s = \lfloor\log_3\frac{\pi}{4\tau}\rfloor$.
	It can quickly verified that $3^s\tau \in (\frac{\pi}{12},
\frac{\pi}{4}]$.

\begin{figure}[H]
    \centering
    \includegraphics[width=0.5\linewidth]{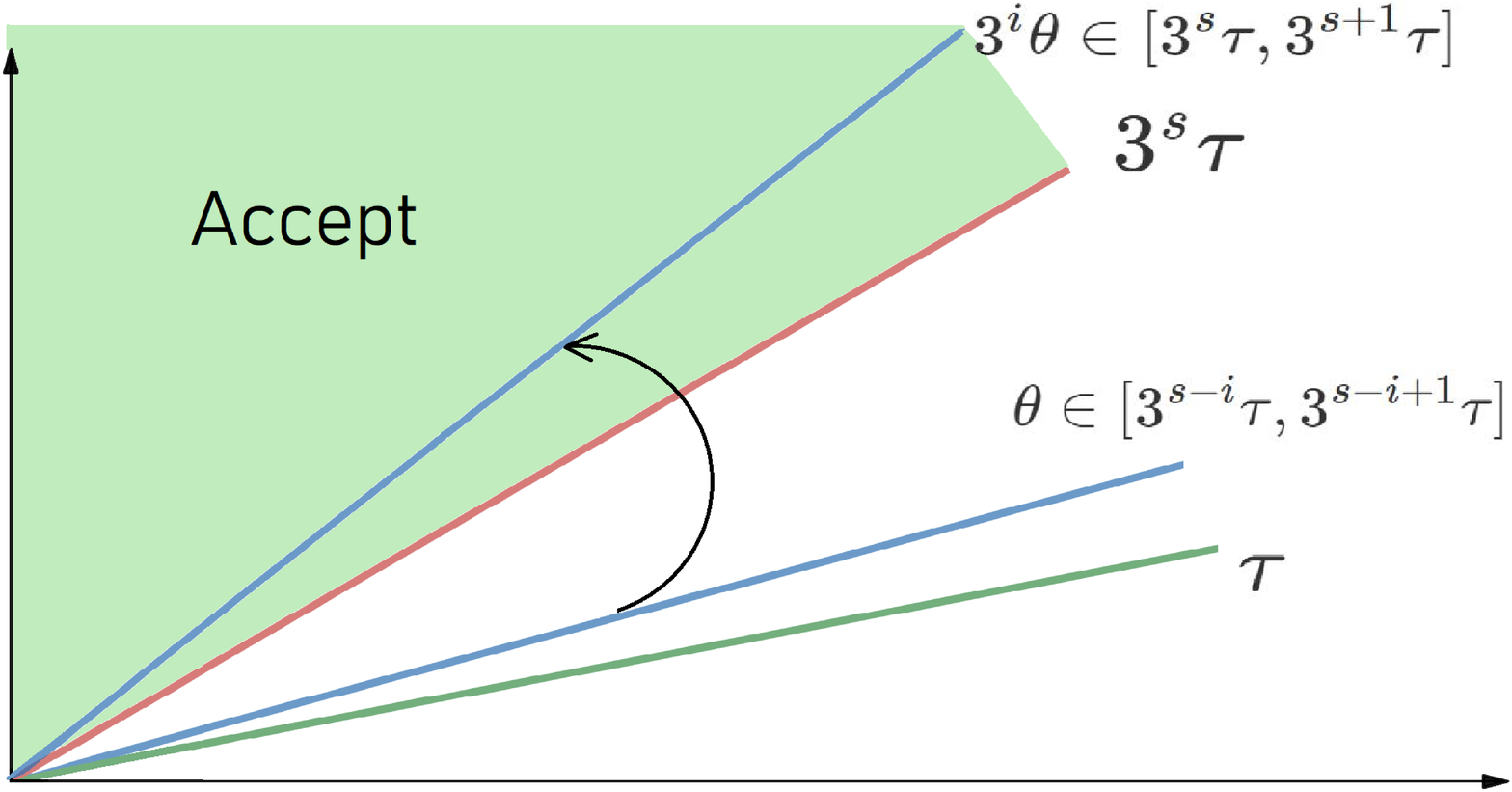}%
    \includegraphics[width=0.5\linewidth]{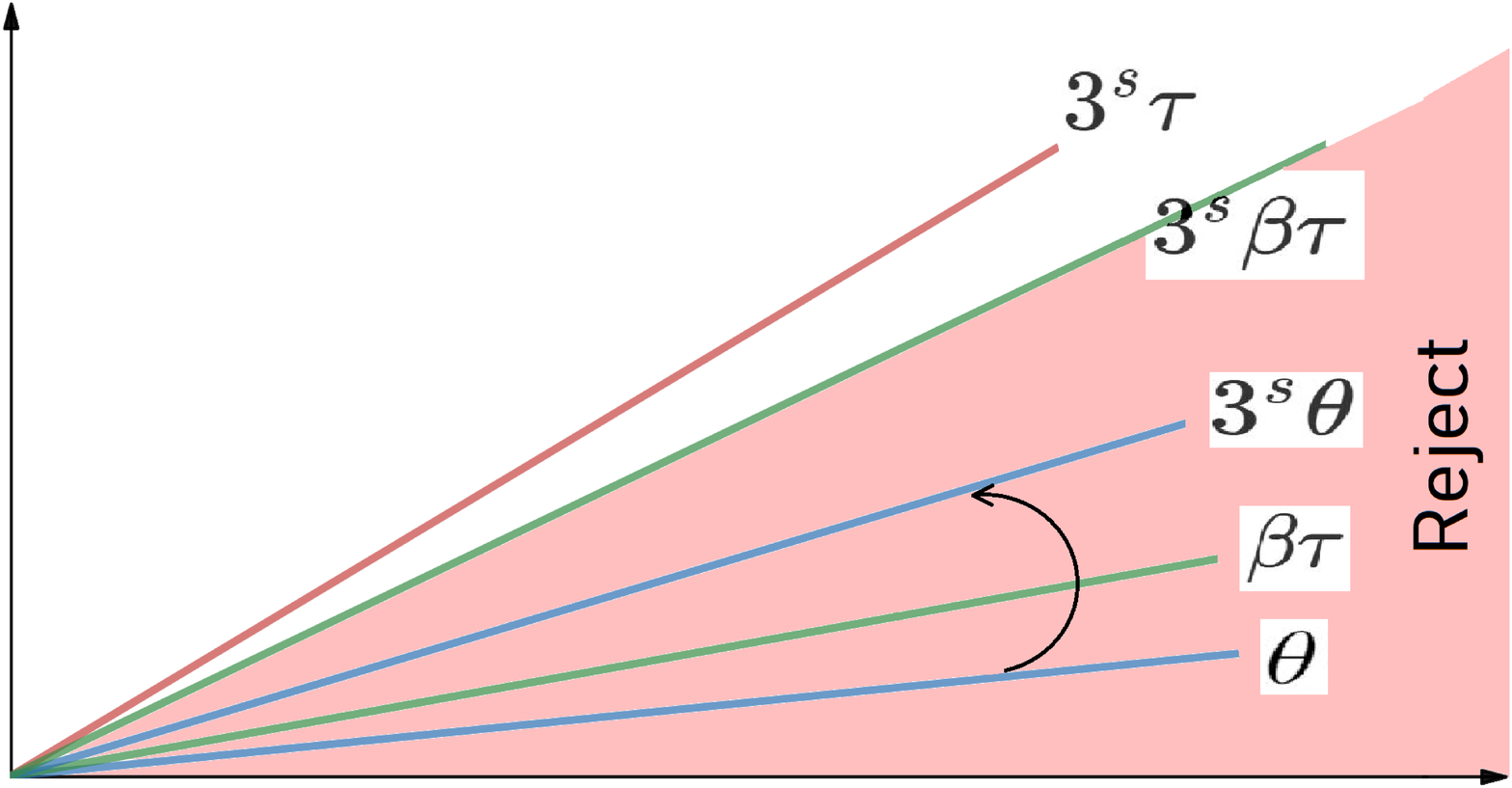}
    \caption{Case ``$\sin \theta \ge \sin\tau$'' (left) and case ``$\sin \theta
    \le \beta\sin\tau$'' (right) of Algorithm~\ref{algo:decision_tau}, before
    and after amplification\label{fig:lemma-cases}}
\end{figure}

	First, consider the case of $\sin \theta \ge \sin \tau$ that is
equivalent to $\theta \ge \tau$ (refer to Figure~\ref{fig:lemma-cases}). 
	Notice that for any $\theta \in [\tau, \frac{\pi}{2}]$, there exists
some $R_i$ such that $\theta \in R_i$. Consider the $i$-th iteration in the
Algorithm in which we set $k_i = \frac{1}{2}(3^i-1)$. For any $\theta \in
R_{i\ne0}$, $(2k_i+1)\theta = 3^i\theta \in [3^s\tau,
3\cdot3^s\tau]\subseteq[3^s\tau, 3\pi/4]$ and for $\theta \in R_0$,
$(2k_i+1)\theta \in [\frac{\pi}{12}, \frac{\pi}{2}] \subseteq [3^s\tau,
3\pi/4]$. Since $3^s \tau \in (\frac{\pi}{12}, \frac{\pi}{4}]$, therefore,
after amplification 
$\sin^2[(2k_i +1)\theta] \ge \sin^2 3^s\tau$. So, the probability $p$ of
observing the output qubit in $\ket{1}$ satisfies $p \ge \sin^2 3^s\tau$.
Therefore, using additive amplitude
estimation with $\epsilon'$ and $\delta'$ as specified in the algorithm will
ensure that $\tilde{p} \ge p - \epsilon' \ge \epsilon^*$
holds with probability at least $1-\delta'$. Hence, the probability that the
algorithm with return {\tt accept} in the $i$-th iteration is at least
$1-\delta'$ and the probability that the algorithm will {\em correctly} return {\tt accept}
eventually is also at least $1-\delta' \ge 1-\delta$.


	Next, consider the case where $\sin\theta\le\beta\sin\tau$ (refer to
Figure~\ref{fig:lemma-cases}).
As per the trigonometric claim above, this implies that $\theta \le \beta\tau$.
Therefore, for any $i=1 \ldots s$, $(2k_i+1)\theta\le(2k_i+1)\beta\tau \le
3^s\beta\tau$. This implies that the probability $p$ defined above satisfies $p
\le \sin^2(3^s\beta\tau)$.
Again using the additive amplitude estimation in a similar manner as above will
ensure that $\tilde{p} \le p+\epsilon' \le \epsilon^*$ with probability at least $1-\delta'$. 
Hence, the probability that the algorithm will return {\tt accept} in a specific
iteration is at most $\delta'$. Therefore, the probability that the algorithm
will return {\tt accept} in {\em any} of the $i=0 \ldots s$ iterations is at
most $(1+s)\delta' = \delta$, which means that the probability that the
algorithm will {\em correctly} return {\tt reject} is also at least $1-\delta$.

Having shown that Algorithm~\ref{algo:decision_tau} returns the correct answer
to its decision problem with error at most $\delta$, now we explain the query
complexity of the algorithm. We will use $M$ to denote the number of queries
made by the additive amplitude estimation algorithm with parameters $\epsilon'$
and $\delta'$; it was shown in Section~\ref{sec:amp-est} that $M =
O(\frac{\pi}{\epsilon'}\log\frac{1}{\delta'})$.
We first need a lower bound on $\epsilon' = \half(\sin^2 3^s\tau - \sin^2 3^s
\beta\tau)$. Using the fact that $3^s\tau \in (\pifrac{12},\pifrac{4}]$ and the trigonometric facts stated above we derive the following:
\begin{align*}
    \sin^2 3^s\tau - \sin^2 3^s\beta\tau & \ge \sin^2 3^s\tau - \beta \sin^2
3^s\tau \\
    & = (1-\beta)\sin^2 3^s\tau > (1-\beta)\pifrac{12}
\end{align*}

Therefore, hiding all constants in the big-$O$ notation, $M =
\tilde{O}(\frac{1}{(1-\beta)} \log \frac{1}{\delta})$.
Now, in Algorithm~\ref{algo:decision_tau}, we can see that the oracle $\A$ is
called a total of $(1+M)k_i$ times at each iteration as the oracle is explicitly
called $k_i$ times during the amplitude amplification and the amplitude
estimation subroutine itself calls the amplitude amplification $M$ times. So,
the total number of calls to the oracle in the algorithm can be expressed as:
	\begin{align*}
	    & \sum_{i=0}^{s}(1+M)k_i  = \frac{1}{2}(1+M)\sum_{i=0}^{s}(3^i-1) <
	    \frac{1}{2}(1+M) \frac{3\pi}{8\tau}\\
		&
	    =\tilde{O}\left(\frac{1}{(1-\beta)\tau}\log\frac{1}{\delta}\right) =
	    \tilde{O}\left(\frac{1}{(1-\beta)\sqrt{t}}\log\frac{1}{\delta}\right)
	\end{align*}
	where we used $\frac{1}{sin^{-1}\sqrt{t}} < \frac{1}{\sqrt{t}}$ in the last inequality.
	

Suppose $\A$ has bounded errors, say $\rho_n$ and
$\rho_y$; then for ``no''-inputs $p \le \rho_n$ and for ``yes''-inputs,
$p \ge (1-\rho_y)$. Further suppose we want to reduce its error to at most
$\delta < \{ \rho_n, \rho_y \}$. Algorithm~\ref{algo:decision_tau} can be applied to $\A$
by setting parameters $t$ to $1-\rho_y$ and $t'$ to $\rho_n$, and, as
shown above, will return {\tt ``accept''} for ``yes''-inputs, as well as return
{\tt ``reject''} for ``no''-inputs, both with probability at least $1-\delta$.
What we obtain is an algorithm that acts on the same input state as $\A$, and
observed using the same measurement operators, but makes at most $\delta$ error
in identifying ``yes'' and ``no''-inputs.
This is our proposal to reduce the error of $\A$ in a generic manner.
The number of calls that will be made to $\A$ (and $\A^\dagger$) in the reduced
error algorithm will be
at most $O(\frac{1}{\sqrt{1-\rho_y}-\sqrt{\rho_n}} \log
\frac{1}{\delta})$.~\footnote{The exact expression, along with all constants, turns out to be
\begin{align*}
    \half \cdot \frac{3\pi}{8 \sin^{-1}\sqrt{t}}  \left(1 + 7 \left\lceil \frac{36}{1-\beta} \right\rceil
    \textstyle(\ln\frac{1+s}{\delta})^{1/3}\right)\\
    \lessapprox \frac{3\pi}{16\sqrt{1-\rho_y}} + \frac{48\pi}{\sqrt{1-\rho_y} -
    \sqrt{\rho_n}} \textstyle(\ln\frac{1+s}{\delta})^{1/3})
\end{align*}
}

%
%

\section{Weight Decision Algorithm}\label{sec:mwdp}


Given an $n$-bit Boolean function $f()$ and two parameters $0 < k_1 <
k_2 < 2^n$, suppose it is given that
either $wt(f) \le k_1$ or $wt(f) \ge k_2$.
We define the {\em Weight Decision} problem, denoted by $\WDP_{k_1,k_2}$,
as the question of determining whether $wt(f) \le k_1$ or $wt(f) \ge k_2$. The objective
is to minimize the number of calls to $f()$ that is given as input in the usual form of a blackbox operator
$U_f:\ket{x}\ket{b} \mapsto \ket{x}\ket{b \xor f(x)}$ where $x \in \{0,1\}^n, b \in \{0,1\}$.

$\WDP$ is fairly versatile in its applicability to Boolean function problems.
For example, $\EWDP$ is a restricted version of $\WDP$ where it given that either $wt(f) = k_1$
or $wt(f)=k_2$ and the problem is identify which case it is. The decision
version of the unordered ``Grover's'' search problem is to identify whether
$wt(f)=0$ or $wt(f) \ge 1$ which is $\WDP_{0,1}$.
The Deutsch's problem and the Deutsch-Jozsa's problem acts on
Boolean functions that are either constant or balanced and their objective is to
determine which one it is; for $n$-bit functions this is equivalent to identifying whether $wt(f) \in
\{0,2^n\}$ or $wt(f)=2^{n-1}$. Following the technique suggested by
Bera~\cite{bera2015different}, one can define the function $g(x)=f(x) \xor f(0)$;
both the problems can now be reformulated as $\EWDP$ with weights $0$ and
$2^{n-1}$ with the function $g()$
as input.

\begin{figure}[!htb]
    \includegraphics[width=0.8\linewidth]{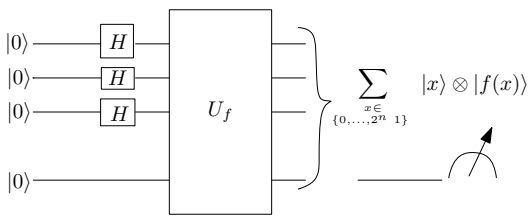}
    \caption{\label{fig:wdp_basic}Quantum circuit for $\WDP$ with bounded error}
\end{figure}

There is a very simple quantum algorithm for
$\WDP_{k_1,k_2}$, illustrated in Figure~\ref{fig:wdp_basic}.
For ease of explanation, we recast the problem as a decision problem --- we denote
functions for which $wt(f) \ge k_2$ as ``yes''-inputs and functions for which
$wt(f) \le k_1$ as ``no''-inputs.
Consider the algorithm that first runs the above circuit and then outputs ``yes'' (i.e., claims that the function
satisfies $wt(f) \ge k_2$) if the last qubit is observed in the state $\ket{1}$
upon measurement and outputs ``no'' otherwise. If the input is a ``yes''-input,
then the probability of error is at most $\rho_y = (1-k_2)/2^n$ and if the input
is a ``no''-input, then the probability of error is at most $\rho_n = k_1/2^n$.
These errors can be reduced to any $\delta$ by using the above algorithm
(in Figure~\ref{fig:wdp_basic}) as $\A$ 
in Algorithm~\ref{algo:decision_tau}.
The number of calls to $\A$, and so to $f()$, would be
$O(\frac{\sqrt{2^n}}{\sqrt{k_2} - \sqrt{k_1}})$ --- this is
asymptotically optimal in $n$ for constant  $k_1$ and $k_2$ due to the fact that
$\WDP$ generalizes the unordered search problem which has a $\Omega(\sqrt{2^n})$ lower
bound.

\begin{algorithm}[H]
    \caption{$\MWDP$($f$,$[w_1, w_2, \ldots, w_j]$) \label{algo:mwdp}}
	{\bf Require:} $0 < w_1 < w_2 < \ldots < w_j < 2^n$\\
	{\bf Global parameter:} $\delta$ (error), $k$ (number of possible
    weights)
    \begin{algorithmic}[1]
	\IF{$j==1$}
	    \RETURN ~$w_k$
	\ELSE
	    \STATE $m = \lfloor j/2 \rfloor$, $t = w_{m+1}/2^n$, $t' = w_m/2^n$, $\delta' =
	\delta/\log_2(k)$
	    \STATE $\A$ : quantum circuit for $\WDP$
	(Figure~\ref{fig:wdp_basic}) using $f()$
	    \STATE /* Determine if $wt(f) \le w_m$ or
	$\ge w_{m+1}$ */
	    \IF{$\AS(\A,t,t',\delta')$ accepts}
		\STATE $\MWDP$($f$, $[w_{m+1}, \ldots, w_j]$)
	    \ELSE
		\STATE $\MWDP$($f$, $[w_1, \ldots, w_{m}]$)
	    \ENDIF
	\ENDIF
    \end{algorithmic}
\end{algorithm}

A similar idea can be used to design an algorithm for the $\MWDP$ problem with
$k$ possible weights $\{0 < w_1 < w_2 < \ldots < w_k \}$.
Our bounded-error algorithm for determining $wt(f)$ is described in Algorithm~\ref{algo:mwdp}.
The algorithm recursively searches for the correct weight in the
list $L$ that it maintains. In each recursive call, it uses $\AS$ to determine
if $wt(f)$ lies in the lower half of the weights in $L$ or in the upper half,
and accordingly, discards half of the possible weights from $L$. Specifically,
if $wt(f) \le w_m$, then $\A$'s probability of success is at most $w_m/2^n$ and
otherwise, it is at least $w_{m+1}/2^n$; therefore, $t$ and $t'$ are set to
$w_{m+1}/2^n$ and $w_m/2^n$, respectively.
The algorithm makes an error if and only if any of the $\AS$
makes an error, and since there are $\log_2(k)$ such calls, the maximum error
that Algorithm~\ref{algo:mwdp} can make is $\log_2(k) \cdot \delta' = \delta$.

The trivial classical complexity of exact $\MWDP$ (without any error) with $k$ possible weights is $O(2^n)$.
The best known quantum method for exact $\MWDP$ was also proposed by Choi et
al.~\cite{Choi11} in which the authors made $k-1$ calls to $\EWDP$. Since the
optimal query complexity of $\EWDP$ is $\Theta(\sqrt{2^n})$, therefore,
their approach yields a better-than-classical approach only when $k \ll
\sqrt{2^n}$. Compared to those, our approach has a complexity $\tilde{O}(\sqrt{2^n}
\log_2 k \log\frac{1}{\delta})$ that we next explain, and suffers from a
negligible probability of error $\delta$ --- the dependency of the complexity on $\delta$ being
logarithmic, it is possible to set a very low $\delta$ without heavy increase in the
complexity. Recall that $\MWDP(f,[w_1, w_2, \ldots, w_k])$ makes altogether
$\log_2(k)$ calls to $\AS$ in a recursive manner. When $\AS$ is called with
parameters $t'=w_m/2^n$ and $t'=w_{m+1}/2^n$, the number of calls to $f()$ is at
most $O(\frac{\sqrt{2^n}}{\sqrt{w_{m+1} - \sqrt{w_m}}}\log\frac{1}{\delta'}) =
\tilde{O}(\sqrt{2^n}\log\frac{1}{\delta})$ leading us to the complexity stated before.
In particular, when $k = \Theta(n)$, existing quantum algorithms have the same asymptotic
complexity of $O(2^n)$ as classical algorithms but our approach uses only
$O(n\sqrt{2^n})$ calls to $f()$.

\section{Conclusion}\label{sec:conclusion}
In this note we have described a technique to reduce error in quantum
algorithms in a blackbox manner, akin to the classical approaches of running an
algorithm multiple times. We showed how to use our approach for designing an
efficient low-error algorithm for the Multiple Weight Decision problem.
At the core of our approach is a new quantum algorithm that decides if the
probability of success of an algorithm is less than $p_1$ or more than $p_2$ for
given $p_1 < p_2$. It would be interesting and beneficial to solve its
multi-class version, i.e., given possible ranges, $[0,p_1], (p_1,p_2], \ldots, (p_k,1]$, 
determine the correct range of the success probability.


\begin{acknowledgments}
Second author would like to thank Indraprastha Institute of Information
Technology Delhi (IIIT-Delhi) for hosting him during which this work
was accomplished. 
\end{acknowledgments}

\bibliography{quantum,my-citations}

\appendix
\section{Proof of trigonometric facts\label{appen:lemma-proof}}

\begin{figure}[H]
    \centering
\includegraphics[width=0.6\linewidth]{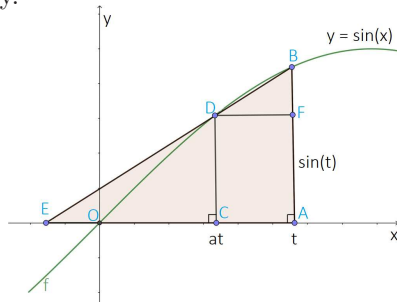}
    \caption{Proof of the fact that $\sin\theta \le a \sin t \implies \theta \le
    at$\label{fig:trig-lemma-proof}}
\end{figure}

We include a quick geometric proof of the trigonometric identity that for any $a
< 1$ and $t \le \pi/2 $, $$\sin\theta \le a \sin t \mbox{ implies that } \theta
\le at.$$
For this consider the right-angled triangles $ABE$ and $CDE$ in
Figure~\ref{fig:trig-lemma-proof}. $E$ is the point where the line segment $BD$
intersects the X-axis and $B$ and $D$ are points on the $sin(x)$ curve
corresponding to $x=t$ and $x=at$, respectively.

We know from geometry that $CDE$ is similar to $ABE$, that is,
$\frac{\sin at}{\sin t} = \frac{CD}{AB} = \frac{EC}{EA}$. From the figure, $EC =
EO + at$ and $EA=EO+t$ which implies that $\frac{\sin at}{\sin t} = \frac{EO +
at}{EO+t} \ge a$. Therefore, $a \sin t \le
\sin at$. Furthermore, we are given that $\sin \theta \le a \sin t$. Combining
the last two facts we get that $\sin \theta \le \sin at$ which in turn implies
that $\theta \le a t$ settling the fact.

In our analysis we make use of the fact that ${\sin at \ge a \sin t}$ for
$a \in (0,1)$ and $t \in [0,\pi/2]$ which follows from the above result.

We make use of another fact which states that $\sin at \le \sqrt{a}
\sin t$ for $t \in [0,\pifrac{4}]$ and $a < 1$ whose proof we discuss now.
Consider the real-valued continuous function $f(t) = a\sin^2 t -
\sin^2 at$. We will now show that $f(t)$ is non-negative for
$t \in [0,\pifrac{4}]$. For showing this, first observe that $f(0) = 0$.
Furthermore, the first derivative satisfies $f'(t) = a(\sin
2t - \sin 2at) \ge 0$ since $a \in (0,1)$ and $t \in
[0,\pifrac{4}]$. This shows that for the specified values of
$t$, $f(t) \ge 0$, or equivalently, $\sin at \le \sqrt{a}
\sin t$.

\end{document}